\documentclass[aps,prl,twocolumn,showpacs]{revtex4}
\usepackage{graphics}

\newcommand{\nl}{\nonumber \\}

\newcommand{\be}{\begin{equation}}
\newcommand{\ee}{\end{equation}}
\newcommand{\bea}{\begin{eqnarray}}
\newcommand{\eea}{\end{eqnarray}}

\newcommand{\Eq}[1]{Eq.\,(\ref{#1})}

\begin{document}

\title{ Incoherent coincidence imaging and its applicability in 
X-ray diffraction
}

\author{Jing Cheng, Shensheng Han}

\affiliation{ Key Laboratory for Quantum Optics 
and Center for Cold Atom Physics, 
Shanghai Institute of Optics and Fine Mechanics, 
Chinese Academy of Sciences, Shanghai, 201800, China
}

\begin{abstract}
Entangled-photon coincidence imaging is a method to nonlocally image an 
object by transmitting a pair of entangled photons through the object 
and a reference optical system respectively. The image of the object 
can be extracted from the coincidence rate of these two photons. 
From a classical perspective, the
image is proportional to the fourth-order correlation function of the
wave field. Using classical statistical optics, we study a particular
aspect of coincidence imaging with incoherent sources.
As an application, we give a proposal to realize 
lensless Fourier-transform imaging, and discuss its applicability 
in X-ray diffraction.
\end{abstract}

\pacs{42.30.Va, 61.10.Dp, 42.25.Kb, 42.50.Ar}

\maketitle

Optical imaging techniques using classic light sources have been 
the primary tools for scientific research and industrial applications.
In recent years, there has been an increasing interest in the field
of quantum imaging, in which nonclassical states of light are used as 
light sources \cite{epjdsyh,rev,jetp,init,
entangle,theory,retrodic,macro,qlith,qoct,qholo,amp}. 
Special attentions are focused on 
entangled-photon coincidence imaging \cite{epjdsyh,rev,jetp,init,
entangle,theory,retrodic,macro}.

The role of entanglement in coincidence imaging leads to some debates 
now. The authors in Ref \cite{entangle} stated 
that only quantum entangled sources can be used to realize 
coincidence imaging, and using classical light sources can not produce 
the image of the object. However, using classically correlated 
beams, the experiment performed in \cite{classic} also produced 
a coincidence image. Moreover, in a recent preprint \cite{bsincoh}, 
it was shown
using quantum theory that an object can be imaged via coincidence
imaging with split incoherent thermal radiation.
In this letter, we give a completely classical 
description of coincidence imaging and obtain a relationship between the 
intensity correlation in the detectors and the source.   
Especially, with a proper choice of the imaging geometry, 
we find it is possible to realize a kind of lensless
Fourier-transform imaging by using an incoherent light source, which
may be applicable for X-ray diffraction. 

\begin{figure}[ht]
{\scalebox{.9}{\includegraphics*{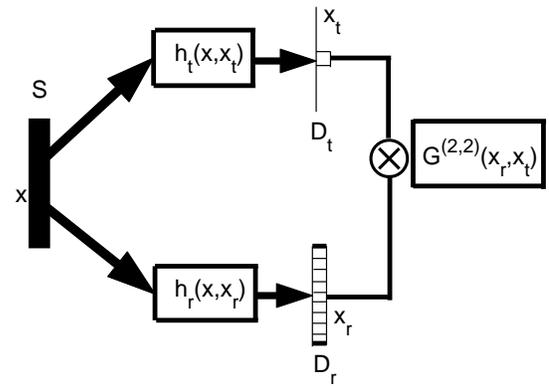}}}
\caption{A setup of entangled-photon coincidence imaging.
The source $S$ emits pairs of entangled photons. One of the photons 
transmits through the test system $h_t(x,x_t)$ which contains an unknown 
object, and the other photon transmits through a known 
reference system $h_r(x,x_r)$. 
Two detectors $D_t$ and $D_r$ record the intensity 
distribution. The coincidence rate $G^{(2,2)}(x_r,x_t)$ 
is measured to give a image of the object.
}
\label{setup}
\end{figure}

An example of the setup of a coincidence imaging system is shown in 
Fig (\ref{setup}) \cite{theory,retrodic}. 
If the source $S$ produces pairs of entangled 
photons, the produced photons are transmitted through a known (reference) 
optical system and an unknown optical system (test) which contains the 
object to be imaged. These two optical systems are characterized 
by their impulse response function $h_r(x,x_r)$ and $h_t(x,x_t)$ 
respectively. Two detectors $D_1$ and $D_2$ record the intensity 
distribution of the test and reference photons. The coincidence 
rate of photon pairs at these 
two detectors ($G^{(2,2)}(x_r,x_t)$) is proportional to 
the fourth-order correlation function of the optical fields,
\be
G^{(2,2)}(x_r,x_t)=<I(x_r)I(x_t)>,
\ee
where $<...>$ means the ensemble average. 
In entangled-photon imaging, 
the object can be extracted from the marginal coincidence rate 
($I^{(2)}(x_r)=\int\!dx_t\, G^{(2,2)}(x_r,x_t)$) or the conditional coincidence 
rate ($I_0^{(2)}(x_r)=G^{(2,2)}(x_r,0)$) \cite{theory}.
Although the reference photons do not pass through the object, the 
object contained in the test system can be 
imaged at the reference detector.
Such a nonlocal imaging technique may be useful for secure information 
transfer.

Now, suppose the light source $S$ is a classical light source, we use 
classically statistical optics to describe the coincidence imaging 
process. In the framework of fluctuating optical fields \cite{goodman}, 
the forth-order correlation function $G^{(2,2)}(x_r,x_t)$ relates to the 
optical fields in the reference and test detectors by
\be
G^{(2,2)}(x_r,x_t)=<E_r^{*}(x_r)E_t^{*}(x_t)E_r(x_r)E_t(x_t)>,
\label{G22}
\ee
where $E_r(x_r)$ is the optical field in the reference detector and 
$E_t(x_t)$ is the optical field in the test detector. For simplicity, 
we assume the source is quasi-monochromatic with a mean wavelength 
$\lambda$. Only one transverse dimension ($x$) is considered though 
the generalization to two transverse dimension is straightforward.

If the optical field in the source is represented by $E(x)$, the 
propagation of $E(x)$ through two different optical systems leads to
\be
E_k(x_k)=\int\!dx\, E(x)h_k(x,x_k) ,
\label{Ek} 
\ee
where $k=r,t$.
Note that $h_r$, $h_t$ are deterministic functions, 
by substituting \Eq{Ek} into \Eq{G22}, we have
\bea
&& G^{(2,2)}(x_r,x_t)  \nl
&=&\int\!dx_1dx_1'dx_2dx_2'\, 
G^{(2,2)}(x_1,x_1',x_2,x_2')     \nl
&& \times h_r(x_1,x_r)h_r^{*}(x_1',x_r)
h_t(x_2,x_t)h_t^{*}(x_2',x_t) .
\label{prop}
\eea
where 
\be
G^{(2,2)}(x_1,x_1',x_2,x_2')=
<E^{*}(x_1')E^{*}(x_2')E(x_1)E(x_2)>
\ee
is the fourth-order correlation function of the optical 
fields at the light source.

\Eq{prop} establishes the relation between the coincidence rate at 
the detectors and the correlation at the source. We need 
to know the properties of the light source to go further. 
In many cases, the field fluctuations of a classical 
light source can be modeled by a complex circular Gaussian 
random process with zero mean \cite{goodman}, then 
\bea
G^{(2,2)}(x_1,x_1',x_2,x_2')&=& G^{(1,1)}(x_1,x_1')G^{(1,1)}(x_2,x_2')+
\nl
     && G^{(1,1)}(x_1,x_2')G^{(1,1)}(x_2,x_1'),
\label{expan}
\eea
where $G^{(1,1)}(x_i,x_j)$ is the second-order correlation function 
of the fluctuating source field, represented by 
$G^{(1,1)}(x_i,x_j)=<E^{*}(x_i)E(x_j)>$, and satisfies 
$G^{(1,1)}(x_i,x_j)=\left [ G^{(1,1)}(x_j,x_i) \right ]^{*}$.

Substituting \Eq{expan} into \Eq{prop}, we get
\begin{widetext}
\bea
G^{(2,2)}(x_r,x_t)
  &=&
\left(
\int\!dx_1dx_1'\, G^{(1,1)}(x_1,x_1')h_r(x_1,x_r)h_r^{*}(x_1',x_r)
\right) \times
\left(
\int\!dx_2dx_2'\, G^{(1,1)}(x_2,x_2')h_t(x_2,x_t)h_t^{*}(x_2',x_t)
\right)  \nl
  && +
\left(
\int\!dx_1dx_2'\, G^{(1,1)}(x_1,x_2')h_r(x_1,x_r)h_t^{*}(x_2',x_t)
\right) \times
\left(
\int\!dx_2dx_1'\, G^{(1,1)}(x_2,x_1')h_t(x_2,x_t)h_r^{*}(x_1',x_r)
\right)  \nl
  &=& <I_r(x_r)><I_t(x_t)>
+ \left | 
\int\!dx_1dx_2'\, G^{(1,1)}(x_1,x_2')h_r(x_1,x_r)h_t^{*}(x_2',x_t)
\right |^2  ,
\label{main}
\eea
\end{widetext}

The first term in the right side of \Eq{main} is the multiplication of 
the intensity distribution at the reference and test detectors, and 
cannot be used to realize the coincidence imaging 
\cite{entangle,theory}. However, if 
$G^{(1,1)}(x_1,x_2)$ is not factorable, the second term in the 
right side of \Eq{main} has the similar form as in the entangled-photon 
coincidence imaging, apart from the presence of a phase conjugated $h_t^{*}$ 
instead of $h_t$. Since a second-order correlation function of 
a classical light source is factorable only when the source is fully 
coherent, we can perform the coincidence imaging using partially 
coherent or incoherent light sources.

Let us introduce the intensity fluctuations in the two detectors
\be
\Delta I_k(x_k)=I_k(x_k)-<I_k(x_k)>,
\ee
in which $k=r,t$. The correlation between the intensity fluctuations 
at the reference and test detectors is 
\bea
&& <\Delta I_r(x_r) \Delta I_t(x_t)> \nl
&=& \left | 
\int\!dx_1dx_2'\, G^{(1,1)}(x_1,x_2')h_r(x_1,x_r)h_t^{*}(x_2',x_t)
\label{DII}
\right |^2 .
\eea
This correlation function is experimentally measurable. A similar 
result has been derived in Ref. \cite{bsincoh}, but quantum theory 
is used in the derivation.

Based on \Eq{DII}, we propose a scheme to realize lensless 
Fourier-transform imaging by selecting proper $h_r$ and $h_t$. 

Suppose the light source is fully spatially incoherent, then 
\be
G^{(1,1)}(x_1,x_2)=I(x_1)\delta(x_1-x_2),
\label{inG11}
\ee
where $I(x)$ is the intensity distribution of the source 
and $\delta(x)$ is the Dirac delta function.

Further, the reference system contains nothing but free-space 
propagation from $S$ to $D_r$.
Under the paraxial approximation,  
the impulse response function of the reference system is
\be
h_r(x,x_r)=\frac{e^{-ikd_r}}{i\lambda d_r}
 \exp\left\{\frac{-i\pi}{\lambda d_r}(x-x_r)^2 \right\} .
\label{inhr}
\ee
where $\lambda$ is the source wavelength and $k=2\pi/\lambda$ 
is the wave number, $d_r$ is the distance between $S$ and $D_r$. 

The test system comprises an object at a distance $d_1$ from 
$S$ and a distance $d_2$ from $D_t$. 
The wave emitted from the source propagates freely to the object 
characterized by the transmittance $t(x')$, then after transmission, 
it propagates freely to the test 
detector. The impulse response function of such a test system is
\begin{widetext}
\be
h_t(x,x_t)= \int\!dx'\, 
\frac{e^{-ikd_1}}{i\lambda d_1}
 \exp\left\{\frac{-i\pi}{\lambda d_1}(x-x')^2 \right\} 
  t(x') \frac{e^{-ikd_2}}{i\lambda d_2}
 \exp\left\{\frac{-i\pi}{\lambda d_2}(x_t-x')^2 \right\} .
\label{inht}
\ee

Substituting Eqs. (\ref{inG11},\ref{inhr},\ref{inht}) into 
\Eq{DII}, after some calculations, we have
\bea
&& <\Delta I_r(x_r) \Delta I_t(x_t)> \nl
&=& \left | 
\int\!dx'dx\, 
I(x) \frac{e^{ikd_r}}{-i\lambda d_r}
 \exp\left\{\frac{i\pi}{\lambda d_r}(x-x_r)^2 \right\}
\frac{e^{-ikd_1}}{i\lambda d_1}
 \exp\left\{\frac{-i\pi}{\lambda d_1}(x-x')^2 \right\} 
t(x') \frac{e^{-ikd_2}}{i\lambda d_2}
 \exp\left\{\frac{-i\pi}{\lambda d_2}(x_t-x')^2 \right\} 
\right |^2  .
\label{xxxx}
\eea
If the source is large enough and the intensity distribution is 
uniform, we can regard $I(x)=I_0$, then \Eq{xxxx} becomes 
\be
 <\Delta I_r(x_r) \Delta I_t(x_t)> 
= \left | 
\int\!dx'\, 
I_0 \frac{e^{-ik(d_1-d_r)}}{i\lambda (d_1-d_r)}
 \exp\left\{\frac{-i\pi}{\lambda (d_1-d_r)}(x_r-x')^2 \right\} 
t(x') \frac{e^{-ikd_2}}{i\lambda d_2}
 \exp\left\{\frac{-i\pi}{\lambda d_2}(x_t-x')^2 \right\} 
\right |^2  .
\label{last}
\ee
Selecting $d_1$, $d_2$ and $d_r$ to satisfy $d_1-d_r=-d_2$, the 
quadratic terms of $x'$ can be canceled, \Eq{last} has the form
\bea
<\Delta I_r(x_r) \Delta I_t(x_t)> 
&=& \left | 
\int\!dx'\, 
\frac{I_0}{\lambda^2 d_2^2}
 \exp\left\{\frac{-i\pi}{\lambda d_2}(x_t^2-x_r^2) \right\} 
t(x') 
 \exp\left\{\frac{i2\pi(x_t-x_r)x'}{\lambda d_2} \right\} 
\right |^2  \nl
&=& \frac{I_0^2}{\lambda^4 d_2^4} 
    \left | T \left( \frac{2\pi(x_t-x_r)}{\lambda d_2} 
    \right) \right |^2  ,
\label{end}
\eea
\end{widetext}
where $T(q)$ is the Fourier transformation of $t(x')$.
So the correlation function between the intensity fluctuations at the 
reference and test detectors is the Fourier transformation of 
the transmittance of the object. We note
that the appearance of $h^*_t$ rather than $h_t$ in \Eq{main} allows this
particular result to be obtained without the use of a lens anywhere in
the system.

If we measure the conditional correlation function of the 
intensity fluctuations by using a point-like 
test detector located at $x_t=0$,
\be
\Delta I_0^{(2)}(x_r)=<\Delta I_r(x_r) \Delta I_t(0)>  ,
\label{ciref}
\ee
it will generate a image recorded in the reference detector but contains 
the information of the object. Equations (\ref{end}) and 
(\ref{ciref}) then yield 
\be
\Delta I_0^{(2)}(x_r)=\frac{I_0^2}{\lambda^4 d_2^4} 
    \left | T \left( \frac{-2\pi x_r}{\lambda d_2} 
    \right) \right |^2  .
\label{lensless}
\ee
So we found that,
under the conditions of a large, uniform, fully incoherent light source, 
without any optical instruments (such as lens) in the reference and test 
systems, using a point-like detector $D_t$ and 
an array of pixel detectors $D_r$,  
such a coincidence imaging system realizes the function of 
Fourier-transform imaging. 

Due to the success of the
oversampling approach, coherent X-ray diffraction imaging has attracted
much attention recently [16-18]. However, several factors still limit
the imaging quality. Because it is very difficult to fabricate optical
components (such as lenses) that function in the X-ray regime,
free-space propagation is used to obtain the diffraction pattern. Also,
it is well known that currently used X-ray sources are generally
incoherent. To achieve the spatial coherence needed to form high-quality
diffraction patterns, such X-ray sources must be small and far from the object
[19]. These requirements decrease the illumination efficiency and
necessitate the use of high brightness sources such as synchrotron
sources.

The lensless Fourier-transform imaging proposal 
given in this letter can overcome these 
difficulties. In fact, the image obtained in 
\Eq{lensless} is exactly the diffraction intensity pattern of the object. 
Since there is no requirement on the fully coherence, any 
kind of X-ray source can be used to realize X-ray diffraction imaging. 
As our method is insensitive to phase
fluctuations of the source, the signal-to-noise ratio will be better
than that achieved in direct diffraction imaging with an incoherent (or
perhaps even partially coherent) source.
So the incoherent coincidence imaging technique is applicable for 
X-ray diffraction.

Finally, we would like to discuss the effects of the time response 
of the detectors on our new imaging scheme. Generally, the intensity 
correlation $<I_r(x_r) I_t(x_t)>$ is not exactly measurable due to 
the finite time response of the detectors. Instead, we can only 
measure
\bea
&& <I'_r(x_r,t) I'_t(x_t,t+\tau)> \nl
&=& \eta \int_{t-T_R/2}^{t+T_R/2} 
       \int_{t+\tau-T_R/2}^{t+\tau+T_R/2}
      <I_r(x_r,t') I_t(x_t,t'')>  \nl
&&  \times dt'dt''       .
\label{detector}
\eea
where we write down the time dependence explicitly. In \Eq{detector},
$\eta$ is a coefficient and $T_R$ is the average time response of 
the detectors. Since in our imaging scheme, the free space 
propagation distance in $h_r$ and $h_t$ are equal, the time delay 
$\tau \approx 0$. In X-ray range, $T_R$ is much larger than the 
coherent time of the fluctuated fields, so the integration of 
\Eq{detector} will be proportional to the equal-time intensity 
correlation $<I_r(x_r) I_t(x_t)>$ \cite{tai}.
Actually, in recent synchrotron radiation experiments, spatial 
intensity correlation have been measured by using slowly response 
detectors \cite{SRcorr}. The key point is that, since only spatial 
intensity correlation is concerned in our imaging scheme, using slowly 
response detectors will screen the temporal fluctuation and measure 
the spatial intensity correlation only. 

In conclusion, we have shown that a classically incoherent
light source can be used to realize coincidence imaging based on 
the measurement of the correlation between the intensity fluctuations. 
Our treatments are fully classical and do not use quantum theory. 
As an application, a scheme to realize lensless
Fourier-transform imaging is described, which 
may be very useful in X-ray diffraction imaging. 
These results will be generalized to three dimensional and  
the effects of source distribution or other imperfections will 
be considered in future works.

The authors are grateful to the reviewers for their helpful comments. 
This work was partly supported
by the National Natural Science Foundation of China under
Grant 69978023.

\end{document}